\documentstyle[12pt,epsf]{article}
\def\Zop{{\sf Z\!\!Z}}

\newcommand{\ko}{\kappa}
\newcommand{\kop}{\kappa'}
\newcommand{\CP}{Chan Paton\ }

\newcommand{\eps}{\epsilon}

\newcommand{\II}{{\cal I}}
\newcommand{\D}{{\cal D}}
\newcommand{\C}{{\cal C}}

\newcommand{\MM}{{\cal M}}

\newcommand{\NN}{{\cal N}}
\newcommand{\PP}{{\cal P}}

\newcommand{\wt}{\widetilde}
\newcommand{\wh}{\widehat}

\newcommand{\ba}{{\bf a}}
\newcommand{\bb}{{\bf b}}
\newcommand{\bk}{{\bf k}}
\newcommand{\bm}{{\bf m}}

\newcommand{\bd}{{\bar {\rm D}}}

\newcommand{\be}{\begin{equation}}
\newcommand{\ee}{\end{equation}}
\newcommand{\ben}{\begin{eqnarray}\displaystyle}
\newcommand{\een}{\end{eqnarray}}
\newcommand{\refb}[1]{(\ref{#1})}

\newcommand{\sectiono}[1]{\section{#1}\setcounter{equation}{0}}

\def\thefootnote{\fnsymbol{footnote}}

\begin{document}

\hfill\vbox{\hbox{hep-th/9908060} 
\hbox{DAMTP-1999-105}
\hbox{MRI-PHY/P990824}
}\break

\vskip 2.5cm

\centerline{\large \bf Non-supersymmetric D-Brane Configurations}

\centerline{\large \bf  with
Bose-Fermi Degenerate Open String Spectrum}  
\medskip

\vspace*{6.0ex}

\centerline{\large \rm Matthias R.~Gaberdiel \footnote{E-mail:  
{\tt M.R.Gaberdiel@damtp.cam.ac.uk}}}

\vspace*{1.5ex}

\centerline{\large \it Department of Applied Mathematics and
Theoretical Physics}
\centerline{\large \it University of Cambridge, Silver Street}
\centerline{\large \it Cambridge, CB3 9EW, U.K.}

\vspace*{2.0ex}

\centerline{\large \it and}

\vspace*{2.0ex}

\centerline{\large \rm Ashoke Sen
\footnote{E-mail: {\tt asen@thwgs.cern.ch, sen@mri.ernet.in}}}

\vspace*{1.5ex}

\centerline{\large \it Mehta Research Institute of Mathematics}
 \centerline{\large \it and Mathematical Physics}
\centerline{\large \it  Chhatnag Road, Jhoosi,
Allahabad 211019, INDIA}

\vspace*{4.5ex}

\centerline {\bf Abstract}

The spectrum of open strings on various non-BPS D-brane configurations
in type II string theory on a K3 orbifold is analysed. At a generic
point in the corresponding moduli space the spectrum of open strings
does not have any degeneracy between bosonic and fermionic
states. However, there exist special values for these moduli for
which many non-BPS D-brane configurations have an exactly bose-fermi
degenerate open string spectrum at all mass levels. In this case
the closed string exchange interaction between a pair of such D-brane
configurations vanishes at all distances.

\vfill \eject

\tableofcontents

\baselineskip=18pt

\def\thefootnote{\arabic{footnote}}
\setcounter{footnote}{0}

\sectiono{Introduction and Summary}

BPS D-branes carrying identical charges do not exert any force on each
other, and can be at equilibrium at all distances. This is a
consequence of supersymmetry, and reflects the fact that the 
spectrum of open strings living on the world volume of the system has
exact degeneracy between bosonic and fermionic states at all mass
levels. As a result the partition function of the open strings, which
corresponds to the negative of the interaction energy of the pair of
D-branes, vanishes identically.

If we consider a set of non-BPS D-branes, or a system of BPS D-branes
carrying different sets of charges and/or with different orientations
so that the combined system is not supersymmetric, then in general the
spectrum of open strings will not have exact bose-fermi
degeneracy. The open string partition function, and hence the
interaction energy of the D-branes, is then not zero. In this case the
D-branes exert a force on each other, and the system is not in
equilibrium.   

In this paper we analyse explicitly the partition function of open
strings for various non-supersymmetric systems of D-branes in type
IIA/IIB string theory on an orbifold K3. We find that in general this
partition function does not vanish and that it has non-trivial
dependence on the relative distance between the branes. 
However, we also find that as we vary the moduli of the K3 orbifold,
the open string spectrum can develop exact bose-fermi degeneracy
at some special points in the moduli space. For such configurations
the partition function vanishes identically, and to one loop in open
string theory, the D-branes do not exert any force on each other.
Our examples contain pairs of non-BPS D-branes, as well as a system of
BPS branes carrying different charge quantum numbers or orientations 
so that the combined system is not supersymmetric.

The various examples that we shall examine involve either BPS D-branes
of type IIA/IIB string theories wrapped on non-supersymmetric cycles
of the K3 orbifold \cite{9812031,9901014}, or a system of BPS D-branes
each wrapped on a supersymmetric cycle  \cite{9603167}, but such that
the combined system breaks all supersymmetries.\footnote{A simpler
example of a bose-fermi degenerate spectrum on a non-BPS brane
configuration involving a combination of a 9-brane and a $\bar
5$-brane has been recently discussed in \cite{9908023}.} A BPS D-brane
wrapped on a supersymmetric cycle is obtained by starting with a BPS
D-brane in type IIA/IIB string theory on a torus $T^4$, where an 
{\em even} number of tangential directions of the brane are along the
torus. This configuration is then modded out by a $\Zop_2$
transformation generated by $\II_4$, the transformation that reverses
the sign of all the coordinates of $T^4$ \cite{9603167}. On the other
hand, a BPS D-brane wrapped on a non-supersymmetric cycle of K3 is
obtained by starting with a non-BPS D-brane in type IIA/IIB string
theory on $T^4$, where an {\em odd} number of tangential directions of
the brane are along the torus; this is then modded out by the same
$\Zop_2$ transformation $\II_4$. For our analysis we shall use a
T-dual description which maps the orbifold of type IIA/IIB on $T^4$ by
$\II_4$ to the orbifold of type IIB/IIA on $T^4$ by 
$g=\II_4\cdot (-1)^{F_L}$, where $F_L$ denotes the contribution to the 
space-time fermion number from the left-moving sector of the
world-sheet. In this description, the T-dual of a BPS D-brane wrapped
on a supersymmetric cycle of K3 is obtained by taking a
BPS D-brane on $T^4$ with an {\em odd} number of tangential directions
along the torus and then modding out the theory by $g$. On the other
hand the T-dual of a BPS D-brane wrapped on a non-supersymmetric cycle
of K3 is obtained by taking a non-BPS D-brane with an
{\em even} number of tangential directions along the torus, and then
modding out the theory by $g$. In all the examples that we discuss we
take $T^4$ to be a direct product of four circles with no background
anti-symmetric tensor field.

The first example that we shall consider is that of a single non-BPS
D$p$ brane of IIB/IIA situated at one of the orbifold fixed points of
$T^4/g$, all of whose tangential directions extend along the
non-compact space-time. (In the dual type IIA/IIB theory on K3 this
corresponds to a BPS $(p+2)$-brane wrapped on a non-supersymmetric
2-cycle of K3.)  This system is stable when the radius of each circle 
of the torus is larger than $(1/\sqrt 2)$.\footnote{We are using
$\alpha'=1$ units.} We find that precisely when all the four radii are
at their critical value, the spectrum of open strings living on the
system develops exact bose-fermi degeneracy, and thus the one loop
open string partition function vanishes. This implies that to this
order, the force between a pair of branes of this type due to closed
string exchange vanishes at all distance scales. We also find that
when any of the radii is above the critical radius, the force is
repulsive at all distances.

The second example involves a pair of anti-parallel BPS D$(p+1)$ 
branes of type IIB/IIA string theory on $T^4/g$, where one of the
tangential directions of the branes is along one of the circles of
the torus, and the other tangential directions 
extend along the non-compact space-time.
The two branes could be separated along the non-compact
directions transverse to the brane, but 
both branes lie along a common fixed line of $T^4/g$.
(In the dual type
IIA/IIB theory on K3 orbifold, this represents a pair of D$(p+2)$
branes, each wrapped on a supersymmetric 2-cycle.) This system is
stable if the radius of the circle tangential to the brane is smaller
than $\sqrt 2$, and the radii of the circles transverse to the brane
are each larger than $(1/\sqrt 2)$. Again we find that precisely when
all the radii take their critical values, the open string spectrum on
this system develops exact bose-fermi degeneracy, even though the
brane configuration is not supersymmetric.

The third example is that of a pair of non-BPS D$p$-branes of type
IIB/IIA placed at two different orbifold points of $T^4/g$, $-$ at
diametrically opposite points of one of the circles of $T^4$, $-$ with
all directions tangential to the 
branes lying along the non-compact
directions. This system is stable when the radius of this special
circle is larger than $\sqrt 2$, and the radii of the other circles
are larger than $(1/\sqrt 2)$. Again when all the radii are at their
critical value, the open string spectrum on this system develops exact
bose-fermi degeneracy. At the critical point there is a marginal
deformation which interpolates between this system and the one in
the previous example; but one can show that the bose-fermi degeneracy
does not survive along this line of marginal deformation.

The fourth example is that of a pair of BPS D$(p+1)$ branes in IIB/IIA
on $T^4/g$, each with one tangential direction along a circle of the 
torus, but unlike the second example where these directions are
anti-parallel, we take the two directions to be along two orthogonal
circles of the torus representing fixed lines on $T^4/g$ that intersect
at a fixed point. The two D-branes share the same $p+1$ non-compact
tangential directions. The spectrum of open strings living on this
system develops exact bose-fermi degeneracy when each of the two
circles transverse to both the branes has radius $(1/2)$. This is also
the critical radius below which this system of branes develops a
tachyonic mode and hence becomes unstable.

There are two ways of computing the partition function of open strings
living on the D-brane system, $-$ directly from the spectrum of open
strings, or using the boundary state formalism to represent the
D-branes as a source for closed strings, and then computing the
amplitude for emission and reabsorption of closed strings. In the open
string calculation it is easy to work out the normalisation of the
partition function, but sometimes it is difficult to know the rules
for projection under various symmetries. In the closed string
computation it is easy to find the rules for projection under various
symmetries, but 
the computation of the normalisation factors is difficult. We
use both approaches. In section \ref{s1}\ we use the known spectrum of
open strings on BPS and non-BPS D-branes in type II string theory on
$T^4$ modded out by $g$ to compute the partition function in examples
1 and 2 in the open string formalism. Then in section \ref{s2}\ we
repeat the analysis for these examples in the boundary state
formalism, and in the process fix the overall normalisation of the
boundary states by comparing with the known answers in section
\ref{s1}. Then we carry out the analysis of examples 3 and 4 using the
boundary state formalism.

Theories with accidental bose-fermi degeneracy have been used
before \cite{9807076,9807213,9808056,9808095,9810129,9904092} in
postulating the existence of non-supersymmetric string theories with
vanishing cosmological constant. It is tempting to speculate that the
type of examples described here could be useful in constructing
orientifold duals of these theories.

The existence of non-BPS brane configurations which do not exert any
force on each other also opens up the possibility of putting a large
number of such systems together, unless higher loop corrections
generate a repulsive interaction between these systems. In this case
we might expect the near horizon geometry of this system to be
described reliably by a solution of the supergravity equations of
motion, thereby giving rise to new relations between
non-supersymmetric field theories and string theory in the spirit of
Refs. \cite{9711200,9802109,9802150}. 

\sectiono{Examples of non-BPS D-brane configurations with exact
bose-fermi degeneracy} \label{s1} 

In this section we shall construct examples of stable non-BPS brane
configurations for which the spectrum of open strings living on the
brane has exact bose-fermi degeneracy at all mass levels, and as a
result the force between these branes due to single closed string
exchange vanishes at all distances. The theory that we shall consider
is type IIB/IIA string theory on $T^4$ modded out by 
$\II_4\cdot (-1)^{F_L}$, with $\II_4$ denoting the transformation
that reverses the sign of all the coordinates of the torus and ${F_L}$
denoting the contribution to the space-time fermion number from the
left-moving sector of the world-sheet. (This theory is T-dual to
type IIA/IIB string theory on $T^4/\II_4$.) Let us denote by 
$x^6,\ldots, x^9$ the coordinates on $T^4$, by $R_6,\ldots, R_9$ the
corresponding radii, and by $x^0,\ldots, x^5$ the non-compact
coordinates. We shall assume that we have Dirichlet boundary 
conditions on the brane along the $x^1,x^2$ directions, and use a
light-cone gauge formalism with $x^1,x^2$ as the light-cone 
directions; a standard D-brane (with Neumann boundary conditions in
the time-direction) can be obtained from this by a double Wick
rotation \cite{9604091}.

For a given configuration of D-branes, the object of interest is the
open string partition function:
\be \label{ex1}
Z = \int {dt\over 2t}\, Tr_{NS-R} (e^{- 2 t H_o} \PP)\, ,
\ee
where NS and R denote Neveu-Schwarz and Ramond sectors, respectively,
$\PP$ is an appropriate projection operator, and $H_o$ is the open
string Hamiltonian:
\be \label{ex2}
H_o= \pi {\vec p}^2 + {1\over 4\pi} {\vec w}^2
+ \pi \sum_{\mu =0,3,\ldots 9}  
[\sum_{n=1}^\infty \alpha^\mu_{-n} \alpha^\mu_n + \sum_{r>0} r
\psi^\mu_{-r} \psi^\mu_r] + \pi C_o\, .
\ee
Here $\vec p$ denotes the open string momentum along the 
directions for which the string has Neumann (N) boundary
conditions at both ends, and $\vec w$ denotes the winding
charge along the directions for which both ends obey Dirichlet (D)
boundary conditions. $\alpha^\mu_n$ and $\psi^\mu_r$ denote
respectively the bosonic and fermionic oscillators satisfying the
commutation relations:  
\be \label{ex3}
[\alpha^\mu_m,\alpha^\nu_n]= m\delta^{\mu\nu}\delta_{m+n,0}, \qquad
\{\psi^\mu_r, \psi^\nu_s\}=\delta^{\mu\nu}\delta_{r+s,0}\, .
\ee
For coordinates satisfying the same boundary condition at both ends of
the open string ({\it i.e.} both Neumann (N) or both Dirichlet (D)) 
$n$ always takes integer values, whereas $r$ takes integer (integer + 
${1\over 2}$) values in the R (NS) sector. On the other hand, for
coordinates satisfying different boundary conditions at the two ends
of the open string (one D and one N) $n$ takes integer+${1\over 2}$
values and $r$ takes integer +${1\over 2}$ (integer) values in the R
(NS) sector. The normal ordering constant $C_o$ vanishes in the
R-sector and is equal to $-{1\over 2}+{s\over 8}$ in the NS sector (in
$\alpha'=1$ units) where $s$ denotes the number of coordinates
satisfying D-N boundary conditions. The trace, denoted by $Tr$, is
taken over the full Fock space of the open string, and also includes a
sum (integral) over various momentum and winding numbers, and a sum
over the different Chan Paton sectors. 

\subsection{Example 1: BPS D-brane wrapped on non-supersymmetric cycle
of K3}
\label{s11}

We begin with a single non-BPS D$p$-brane
\cite{9805019,9806155,9808141,9809111} in type IIB/IIA string theory 
on $T^4/(-1)^{F_L}\cdot\II_4$. The D-brane is situated at one of the
fixed points of $T^4/(-1)^{F_L}\cdot\II_4$, and all $p+1$ directions
on the world-volume extend along the non-compact space-time. $p$ is
even for type IIB string theory, and odd for type IIA string
theory. (In the T-dual description this can be regarded as a BPS
D-$(p+2)$-brane in IIA/IIB wrapped on a non-supersymmetric cycle of K3
\cite{9812031,9901014}.) There are two different \CP sectors, $-$
labelled by the $2\times 2$ identity matrix $I$ and the Pauli matrix
$\sigma_1$, $-$, each having its own rule for GSO ($(-1)^F$)
projection, and  $g\equiv \II_4\cdot (-1)^{F_L}$ projection. Thus
$\PP$ in Eq.~\refb{ex1} is given by:  
\be \label{exx1}
\PP = {1 + (-1)^F\over 2} {1 + g\over 2}\, .
\ee
$(-1)^F$ reverses the sign of all the world-sheet fermions, whereas
$g$ reverses the sign of the world-sheet scalar and fermions
associated with the $x^6,\ldots, x^9$ coordinates. In computing the
partition function \refb{ex1} we also need to know how $(-1)^F$ and
$g$ act on the Fock vacuum in the different sectors; this is known 
\cite{9806155,9808141,9904207}. In the NS sector we have
\ben \label{exx2}
I & : &  (-1)^F = -1 \qquad g = 1 \nonumber \\
\sigma_1 & : & (-1)^F = 1 \qquad g = -1 \, .
\een
Thus when we combine the open string spectrum from the two sectors, we
see that there is no net $(-1)^F$ or $g$ projection, as at a given
mass level we have open string states carrying both $(-1)^F$ charges
and both $g$ charges. However, there is a net $(-1)^F\cdot g$
projection, since the Fock vacuum in both sectors is odd under
$(-1)^F\cdot g$, and hence only those open string states, obtained by
oscillators carrying a net $(-1)^F\cdot g$ charge $-1$ acting on the
vacuum, are allowed in the  spectrum. Thus the combined contribution
of all \CP factors to \refb{ex1} from NS sector states can be written
as 
\be \label{exx3}
\int {dt\over 2t} tr_{NS}
\left(e^{- 2 t H_o} {1 + (-1)^F\cdot g\over 2}\right)\, ,
\ee
where $tr$ now denotes a sum over the oscillators in a single Fock
space, an integration over momenta in the non-compact directions, and
a sum over winding numbers in the compact directions, but does not
contain a sum over the different \CP factors.

The situation in the Ramond sector is even simpler. The ground state is
16-fold degenerate due to the 8 fermionic zero modes, and it contains
an equal number of states with charge 1 and $-1$ under each of the three
operators $(-1)^F$, $g$ and $(-1)^F\cdot g$. Thus
$Tr_R(e^{-2tH_o}(-1)^F)$,
$Tr_R(e^{-2tH_o}g)$, and $Tr_R(e^{-2tH_o}(-1)^F\cdot g)$ all vanish,
and when we combine the spectrum from the two \CP sectors, the
contribution to \refb{ex1} can be written as
\be \label{exx4}
\int {dt\over 2t} {1\over 2} \, tr_{R} (e^{- 2 t H_o} )\, .
\ee

We can now evaluate the contribution from each sector separately. We
get
\be \label{ex5}
tr_{NS} (e^{-2t H_o}) = {A\over (2\pi)^{p+1}} (2t)^{-{p+1\over 2}}
{f_3(\tilde q)^8\over f_1(\tilde q)^8}
\left(\prod_{i=6}^9 \sum_{n_i\in \Zop} \tilde q^{2R_i^2 n_i^2}
\right)\, ,
\ee
where $A$ is the (infinite) $(p+1)$-dimensional volume of the brane in
the non-compact directions,
\be \label{ex6}
\tilde q = e^{-\pi t}\, ,
\ee
and $f_i$ are defined in the usual 
manner \cite{POLCAI}: 
\ben \label{ex7}
f_1(\tilde q) & = & \tilde q^{1\over 12} \prod_{n=1}^\infty
( 1 - \tilde q^{2n})\, , \nonumber
\\
f_2(\tilde q) & = & \sqrt 2 \tilde q^{1\over 12}
\prod_{n=1}^\infty ( 1 + \tilde q^{2n})\, ,
\nonumber
\\
f_3(\tilde q) & = & \tilde q^{-{1\over 24}}
\prod_{n=1}^\infty ( 1 + \tilde q^{2n-1})\, ,
\nonumber
\\
f_4(\tilde q) & = & \tilde q^{-{1\over 24}}
\prod_{n=1}^\infty ( 1 - \tilde q^{2n-1})\, .
\een
The origin of the various factors in \refb{ex5} is as follows. The
$A(2\pi)^{-{p+1}}(2t)^{-(p+1)/2}$ factor comes from integration over
the open string momenta along the non-compact directions and the
$f_3(\tilde q)^8/f_1(\tilde q)^8$ factor represents
the contribution of the bosonic
and the fermionic oscillators. The last factor
$\prod_i \sum_{n_i}
\tilde q^{2 R_i^2 n_i^2}$ comes from the sum over open
string winding modes along the compact directions.

Similarly we get
\be \label{ex8}
tr_{NS} (e^{-2t H_o} (-1)^F\cdot g) = - {A\over (2\pi)^{p+1}}
(2t)^{-{p+1\over 2}}
\cdot 4 \cdot {f_3(\tilde q)^4 f_4(\tilde q)^4\over
f_1(\tilde q)^4 f_2(\tilde q)^4}\, .
\ee
The contribution from the momentum integration along the non-compact
directions remains the same but the oscillator contribution changes,
since four of the bosonic oscillators and four of the fermionic
oscillators change sign under $(-1)^F\cdot g$. There is no
contribution from the winding sector, since $(-1)^F\cdot g$ takes a
state with winding charge $\vec w$ to a state with winding charge
$-\vec w$, and hence the contribution from these states vanishes in
the trace. Finally, the overall $-$ sign reflects the fact that the NS
sector ground state is odd under $(-1)^F\cdot g$ (which is the reason
why this model is free of tachyons).

The contribution from the Ramond sector can also be evaluated in a
straightforward manner. We get
\be \label{ex9}
tr_{R} (e^{-2t H_o}) = {A\over (2\pi)^{p+1}} (2t)^{-{p+1\over 2}}
{f_2(\tilde q)^8\over f_1(\tilde q)^8} \left(\prod_{i=6}^9
\sum_{n_i\in Z} \tilde q^{2R_i^2 n_i^2}\right)\, .
\ee
Comparing with Eq.~\refb{ex5} we see that the contribution from
momentum integration along the non-compact directions, and the sum
over winding modes in the compact direction remains the same. The only
change is in the contribution from the fermionic oscillators.

Combining the contribution from all sectors, and using the (abstruse)
identity
\be \label{ex9a}
f_3(\wt q)^8-f_2(\wt q)^8=f_4(\wt q)^8\, , 
\ee
we see that the total partition function is given by:
\be \label{ex10}
Z = {1\over 2} \int {dt\over 2t} {A\over (2\pi)^{p+1}}
(2t)^{-{p+1\over 2}}
\left[ {f_4(\tilde q)^8\over f_1(\tilde q)^8}\left(\prod_{i=6}^9
\sum_{n_i\in \Zop} \tilde q^{2R_i^2 n_i^2}\right)
- 4 \cdot {f_3(\tilde q)^4 f_4(\tilde q)^4\over f_1
(\tilde q)^4 f_2(\tilde q)^4}\right]\, .
\ee

Let us now consider the case where $R_i={1\over \sqrt 2}$ for each
$i$. In this case we get
\be \label{ex11}
\sum_{n_i\in \Zop} \tilde q^{2R_i^2 n_i^2} =
\sum_{n\in \Zop} \tilde q^{n^2}\, .
\ee
Using the sum and the product representation of the Jacobi
$\vartheta$-function $\vartheta_3(0|\tau)$ \cite{EMOT53},
\be \label{ex12}
\vartheta_3(0|\tau) = \sum_{n\in \Zop} \tilde q^{n^2}
= \prod_{n=1}^\infty (1 -
\tilde q^{2n}) (1 + \tilde q^{2n-1})^2
= f_1(\tilde q) f_3^2(\tilde q)\, ,
\ee
where $\tilde q=e^{2\pi i\tau}$, and the identity
\be \label{ex13}
f_4(\tilde q) {1\over \sqrt 2} f_2(\tilde q) f_3(\tilde q) = 1\, ,
\ee
we get
\be \label{ex14}
\sum_{n\in \Zop} \tilde q^{n^2} = \sqrt 2 \,
{f_1(\tilde q) f_3(\tilde q)\over f_2(\tilde q) f_4(\tilde q)}\, .
\ee
Using Eqs. \refb{ex11} and \refb{ex14}, \refb{ex10} then becomes 
\be \label{ex15a}
Z = 0\, .
\ee
Since the integrand of $Z$ vanishes for all $t$, this shows that there
is exact degeneracy between bosonic and fermionic open string states
at all mass level, although the brane is non-BPS. In order to find the
bosonic and the fermionic spectrum separately, we need to evaluate
\refb{ex9} or the sum of \refb{ex5} and \refb{ex8}. This is given by:
\be \label{ex15}
{A\over (2\pi)^{p+1}} (2t)^{-{p+1\over 2}}
4 \cdot {f_2(\tilde q)^4 f_3(\tilde q)^4
\over f_1(\tilde q)^4 f_4(\tilde q)^4}\, .
\ee
This is proportional to the partition function of bosonic (or
fermionic) open string states stretched between a BPS $p$-brane and a
BPS $(p+4)$-brane in type II string theory.

Note that the critical radii where the spectrum of open strings
develops exact bose-fermi degeneracy correspond precisely to
the values below which the non-BPS D-brane becomes unstable against
decay into a pair of BPS branes \cite{9805019}. This is not a
coincidence. For $R_i>{1\over \sqrt 2}$ the massless spectrum contains
four bosonic states from sector $I$, four fermionic states from sector
$I$, and four fermionic states from sector $\sigma_1$. In order to
have bose-fermi degeneracy at the massless level, we need four extra 
massless bosonic states which are provided by some of the modes from
sector $\sigma_1$ becoming massless at the critical radius. As we
decrease any of the $R_i$ below the critical radius, the corresponding
mode becomes tachyonic signalling an instability in the system.

We can use this result to conclude that when
$R_6=R_7=R_8=R_9={1\over\sqrt 2}$, the force between a pair of non-BPS
branes of this kind vanishes at all distances. To see this we note
that if we consider a pair of such branes separated by a distance $r$
in any of the non-compact directions transverse to the brane, then the
partition function of open strings stretched from one of the branes to
another is given by the same expression as \refb{ex10} except for an
overall extra factor of $\tilde q^{r^2/2\pi^2}$ in the integrand, 
reflecting the energy associated with the tension of the open string
stretched over a distance $r$. Thus at the critical radius the
partition function vanishes, reflecting that the potential energy
$V(r)$ between the pair of branes (which is equal to negative of the
partition function) vanishes identically for all $r$.

Since $\sum_{n_i\in \Zop} \tilde q^{2R_i^2 n_i^2}$ is a monotonically
decreasing function of $R_i$ (as $0<\tilde q<1$), we see that for $R_i
>{1\over \sqrt 2}$ the integrand of Eq.~\refb{ex10} is a negative
definite function. Thus $V(r)$ is positive definite. Furthermore since
$V(r)$ only depends on $r$ via $\tilde q^{r^2/2\pi^2}$, it follows by
the same argument that $V'(r)$ is negative, and hence that $V(r)$ is a
monotonically decreasing function of $r$. Thus for
$R_i>{1\over\sqrt 2}$, where the non-BPS brane is stable, the
interaction between a pair of such branes is repulsive at all
distances.

\subsection{Example 2: Pair of BPS D-branes wrapped on supersymmetric
cycles of K3} \label{s12}

In this section we shall study the example of a parallel BPS
D-$(p+1)$-brane $\bd$-$(p+1)$ brane pair, where one direction extends
along $x^9$, and the other $p+1$ directions on the world-volume 
lie along the non-compact space-time directions. As before we are
considering type IIB/IIA string theory on
$T^4$ modded out by $(-1)^{F_L}\cdot\II_4$, where $\II_4$ inverts
$x^6,\ldots, x^9$.  Consistency requires that $p$ is odd for type IIA
string theory, and even for type IIB string theory. We consider 
the situation where we do not have a Wilson line on any of the branes,
and we take them both to be situated at $x^6=x^7=x^8=0$, while they
can be separated by a distance $r$ along any of the non-compact
directions transverse to the brane. There are two different cases,
depending on whether the brane and the anti-brane carry the same or
opposite twisted sector RR charge \cite{9805019}; here we shall only
consider the situation where the twisted sector RR charge is the same
for both branes. (In the T-dual description this corresponds to a pair
of fractional $p$-branes situated at one of the orbifold fixed 
points of $T^4/\II_4$ in type IIA/IIB string theory on $T^4/\II_4$,
carrying the same twisted sector RR charge, but opposite untwisted
sector RR charge. Each of these can be interpreted as a BPS
$(p+2)$-brane wrapped on a supersymmetric cycle \cite{9603167}. Thus
each system is individually BPS, but the combined system is not BPS,
as the RR charges of the two systems are not aligned.)

We shall compute the partition function of open strings living on this
system. Since each of the branes is individually BPS, the spectrum of
open strings with both ends living on the same brane is automatically
supersymmetric. Thus we only need to focus on the open strings with
one end on the D-brane and the other end on the $\bd$-brane. There are
two such sectors, related by reversal of the orientation of the open
strings. But since both sectors contribute equally to the partition
function we shall restrict our attention to only one sector. Thus
there is no sum over \CP factors, and we can express the partition
function \refb{ex1} as
\be \label{ex16}
Z = \int {dt\over 2t} tr_{NS-R} \left(e^{- 2 t H_o}
{ 1+ (-1)^F\over 2}{1 + g\over 2} \right) \, .
\ee
The action of $(-1)^F$ and $g$ on the world-sheet fields is identical
to that described in the last example, but we need to know the action
on the vacuum. Since the open string is stretched between the D-string
and the $\bd$-string, the NS sector ground state is $(-1)^F$ even. On
the other hand, from the analysis of Ref.~\cite{9805019} we know that
the NS sector ground state carrying zero momentum is projected out
under the $g$ projection. 
Thus it is odd under $g$.\footnote{We are
using a slightly different convention from the one used in
\cite{9805019}. There the action of $g$ and $(-1)^F$ on
the NS sector ground state was defined to be the same as in the case
of open strings with both ends on the D-brane, but the sign of
$(-1)^F$  and $g$ in the projection operator was chosen to be
negative. Here we have chosen the signs of $(-1)^F$ and $g$ in the
projection operator to be positive, but are taking the action of these
operators on the NS sector ground state to be opposite of that for
open strings with both ends on the D-brane.} In the Ramond sector we
have a 16-fold degeneracy, and as in the previous example, the ground
state contains an equal number of states carrying +1 and $-1$
eigenvalues of $(-1)^F$, $g$ and $(-1)^F\cdot g$. Thus only the
$tr_R(e^{-2t H_o})$ term contributes from this sector. 

We can now easily write down the expression for the contribution from
the various sectors, 
\be \label{ex17}
tr_{NS} (e^{-2t H_o}) = {A\over (2\pi)^{p+1}} (2t)^{-{p+1\over 2}}
\tilde q^{r^2/2\pi^2} {f_3(\tilde q)^8\over
f_1(\tilde q)^8}  \Big(\prod_{i=6}^8
\sum_{n_i\in \Zop} \tilde q^{2R_i^2 n_i^2}\Big)
\Big(\sum_{m_9\in \Zop} \tilde q^{2 m_9^2/R_9^2}\Big)\, .
\ee
Compared to \refb{ex5}, there is an extra factor of 
$\tilde q^{r^2/2\pi^2}$ reflecting the effect of the transverse 
separation between the two branes, and the sum over winding
numbers along the 9th direction has been replaced by a sum over 
momenta, since we now have a Neumann boundary condition along this 
direction. Similarly we get
\be \label{ex18}
tr_{NS} (e^{-2t H_o} (-1)^F) = {A\over (2\pi)^{p+1}}
(2t)^{-{p+1\over 2}} \tilde q^{r^2/2\pi^2}
{f_4(\tilde q)^8\over f_1(\tilde q)^8} \Big(\prod_{i=6}^8
\sum_{n_i\in \Zop} \tilde q^{2R_i^2 n_i^2}\Big)
\Big(\sum_{m_9\in \Zop} \tilde q^{2 m_9^2/R_9^2}\Big)\, .
\ee
\be \label{ex19}
tr_{NS} (e^{-2t H_o} g) = - {A\over (2\pi)^{p+1}}
(2t)^{-{p+1\over 2}} \tilde q^{r^2/2\pi^2}
\cdot 4 \cdot {f_3(\tilde q)^4 f_4(\tilde q)^4
\over f_1(\tilde q)^4 f_2(\tilde q)^4}\, .
\ee
\be \label{ex20}
tr_{NS} (e^{-2t H_o} (-1)^F\cdot g) = - {A\over (2\pi)^{p+1}}
(2t)^{-{p+1\over 2}} \tilde q^{r^2/2\pi^2}
\cdot 4 \cdot {f_3(\tilde q)^4 f_4(\tilde q)^4
\over f_1(\tilde q)^4 f_2(\tilde q)^4}\, .
\ee
\be \label{ex21}
tr_{R} (e^{-2t H_o}) = {A\over (2\pi)^{p+1}} (2t)^{-{p+1\over 2}}
\tilde q^{r^2/2\pi^2} {f_2(\tilde q)^8\over f_1(\tilde q)^8}
\Big(\prod_{i=6}^8
\sum_{n_i\in \Zop} \tilde q^{2R_i^2 n_i^2}\Big)
\Big(\sum_{m_9\in \Zop} \tilde q^{2 m_9^2/R_9^2}\Big)\, .
\ee
Substituting these into \refb{ex16} and using the identity
\refb{ex9a}, $Z$ becomes in this case
\be \label{ex22}
\int{dt\over 2t}{1\over 2} {A\over (2\pi)^{p+1}}
(2t)^{-{p+1\over 2}} \tilde q^{r^2/2\pi^2}
\left[{f_4(\tilde q)^8\over f_1(\tilde q)^8}
\Big(\prod_{i=6}^8
\sum_{n_i\in \Zop} \tilde q^{2R_i^2 n_i^2}\Big)
\Big(\sum_{m_9\in \Zop} \tilde q^{2 m_9^2/R_9^2}\Big)
- 4 \cdot {f_3(\tilde q)^4 f_4(\tilde q)^4
\over f_1(\tilde q)^4 f_2(\tilde q)^4} \right] .
\ee
Now we note that for $R_6=R_7=R_8={1\over \sqrt 2}$ and
$R_9=\sqrt{2}$, 
\be \label{ex22a}
\Big(\prod_{i=6}^8
\sum_{n_i\in \Zop} \tilde q^{2R_i^2 n_i^2}\Big)
\Big(\sum_{m_9\in \Zop} \tilde q^{2 m_9^2/R_9^2}\Big)
= (\sum_{n\in \Zop} \tilde q^{n^2})^4 =
\Big(\sqrt 2 \, {f_1(\tilde q) f_3(\tilde q)\over
f_2(\tilde q) f_4(\tilde q)}\Big)^4\, .
\ee
Substituting this into Eq.~\refb{ex22} we see that
\be \label{ex23}
Z = 0\, .
\ee
Thus the spectrum of open strings has exact bose-fermi degeneracy even
though the brane configuration is not supersymmetric. At this critical
point in the moduli space, the force between the pair of branes
vanishes for all separations.

By repeating the analysis of the previous example we can easily verify
that the partition function in the NS or R sector is again
proportional to that of a BPS D-$p$ $-$ D-$(p+4)$ brane system in
type II string theory. Also, if $R_i > {1\over \sqrt 2}$ for
$6\le i\le 8$, and $R_9<\sqrt 2$, the interaction between the pair of 
branes is repulsive at all distances.

\sectiono{The boundary state approach} \label{s2} 

It is instructive to compare the above derivation of the bose-fermi
degeneracy of the open string spectrum with that using the closed
string theory point of view, describing Dirichlet branes as boundary
states \cite{POLCAI,CLNY,ONOISH,ISHIBASHI,BOUNDARYOLD,BOUNDARY}. We
shall use the conventions of \cite{9805019} in the following. Let us
first treat the two cases that we have analysed above in turn in order
to fix the relevant normalisation constants of the boundary states; we
shall then also construct two new examples.

\subsection{Example 1: BPS D-brane wrapped on non-supersymmetric
cycle of K3} \label{s21} 

This is the same example that we studied in section \ref{s11}, $-$
namely, a non-BPS $p$-brane of IIB/IIA on $T^4/(-1)^{F_L}\cdot\II_4$,
all of whose tangential directions lie along the non-compact
space-time. 
Let us denote by $\C\subset\{0,3,4,5\}$ the set of $p+1$ indices 
for which the Dirichlet brane satisfies Neumann boundary conditions,
and denote by $\D$ the complement of $\C$ in $\{0,3,4,5\}$. We also
denote by $\D_c$ the union of $\D$ with $\{1,2\}$, and by
$\hat\D$ the union of $\D$ with $\{6,7,8,9\}$. The boundary state
that represents the non-BPS $p$-brane is of the
form \cite{9806155}\footnote{We shall be
using the convention that $\wt D$ denotes a non-BPS D-brane, and $D$
denotes a BPS D-brane.}
\ben\label{Dp}
|\wt Dp,\ba, \bb, \eps\rangle & = & {1 \over 2}
\Bigl(|Bp,\ba,\bb,+\rangle_{NSNS;U}
              - |Bp,\ba,\bb,-\rangle_{NSNS;U} \Bigr) \nonumber
\\
& & \quad         + {\eps \over 2}
\Bigl(|Bp,\ba,\bb,+\rangle_{RR;T}
              + |Bp,\ba,\bb,-\rangle_{RR;T} \Bigr)\,,
\een
where the first two (last two) states are coherent states in the
untwisted NSNS (twisted RR) sector. $\ba$ denotes the location of the
brane in the non-compact directions with $a^1=a^2=0$, and $\bb$
denotes the location of the brane in the compact directions. 
Since we always take the brane to lie at one of the fixed points of
$T^4/g$,  $b^i$ can only take the values $0$ or $\pi R_i$ 
($6\le i\le 9$). $\eps$ can take values $\pm 1$, and denotes the sign
of the twisted sector RR charge carried by the brane. Up to an overall
normalisation, the coherent states are (uniquely) characterised by the
conditions  
\be \label{esx1}
\begin{array}{rclll}
X^\mu(\tau=0,\sigma) |Bp,\ba,\bb,\eta\rangle & = & a^\mu \qquad &
\hbox{for}
\qquad & \mu\in\D \\
X^\mu(\tau=0,\sigma) |Bp,\ba,\bb,\eta\rangle & = & b^\mu \qquad &
\hbox{for}
\qquad & 6\le\mu\le 9 \\
\partial_\tau X^\mu(\tau=0,\sigma) |Bp,\ba,\bb,\eta\rangle & = & 0
\qquad & \hbox{for} \qquad & \mu\in\C \\
(\psi^\mu(\tau=0,\sigma)-i\eta\tilde{\psi}^\mu(\tau=0,\sigma))
|Bp,\ba,\bb,\eta\rangle & = & 0 \qquad &
\hbox{for} \qquad & \mu\in\hat\D \\
(\psi^\mu(\tau=0,\sigma)+i\eta\tilde{\psi}^\mu
(\tau=0,\sigma)) 
|Bp,\ba,\bb,\eta\rangle & = & 0 \qquad &
\hbox{for} \qquad & \mu\in\C \\
x^\mu |Bp,\ba,\bb,\eta\rangle & = & 0 \qquad & \hbox{for} \qquad &
\mu=1,2\,.
\end{array}
\ee
Here $X^\mu$ denotes the bosonic coordinate field, $\psi^\mu$ and
$\tilde\psi^\mu$ its right- and left-moving superpartner on the
closed string world-sheet, and $x^\mu$ the zero mode of $X^\mu$.
$\eta$ can take values $\pm 1$.
More explicitly the boundary state $|Bp,\ba,\bb,\eta\rangle$ can be
written as
\be\label{boundaryNSNS}
|Bp,\ba,\bb,\eta\rangle_{NSNS;U} = {\cal N} \int
\left(\prod_{\mu\in\D_c} dk^\mu e^{i k \cdot \ba} \right)
\left(\prod_{i=6}^{9} \sum_{m_i\in\Zop}
e^{i m_i b^i/R_i} \right)  
\wh{|Bp,\bk,\bm,\eta\rangle}_{NSNS;U}\,, 
\ee
and
\be\label{boundaryRR}
|Bp,\ba,\bb,\eta\rangle_{RR;T} = 2 i \widetilde{\cal N} \int
\left(\prod_{\mu\in\D_c}
dk^\mu e^{i \bk \cdot \ba} \right)
\wh{|Bp,\bk,\bb,\eta\rangle}_{RR;T} \,, 
\ee
where $\bk$ denotes the momentum in the non-compact directions, and
$m_i/R_i$ the momentum along the $i$th compact direction.
$\wh{|Bp,\cdots,\eta\rangle}$ 
is the coherent momentum eigenstate
\ben\label{bound}
\wh{|Bp,\cdots,\eta\rangle} 
& = & \exp \left(
\sum_{n=1}^\infty \left[
- {1\over n} \sum_{\mu\in\C} \alpha^\mu_{-n} \tilde\alpha^\mu_{-n}
+ {1\over n} \sum_{\mu\in\hat\D} \alpha^\mu_{-n} \tilde\alpha^\mu_{-n}
\right] \right. \nonumber \\
& & \qquad\qquad\left. + i \eta \sum_{r>0} \left[
- \sum_{\mu\in\C} \psi^\mu_{-r} \tilde\psi^\mu_{-r}
+ \sum_{\mu\in\hat\D} \psi^\mu_{-r}
\tilde\psi^\mu_{-r} 
\right] \right) |\cdots,\eta\rangle^{(0)} \,.
\een
Here $\alpha^\mu_n$, $\tilde\alpha^\mu_n$ are the right- and
left-moving modes of $X^\mu$, and $\psi^\mu_r$ and $\tilde\psi^\mu_r$
are the modes of $\psi^\mu$ and $\tilde\psi^\mu$, respectively. $n$ is
integer in the untwisted NSNS sector, and in the twisted RR sector 
for $\mu=0,3,4,5$, and half-integer in the twisted RR sector for
$\mu=6,7,8,9$. $r$ is half-integer in the untwisted 
NSNS sector, and in the twisted RR sector for $\mu=6,7,8,9$, and
integer in the twisted RR sector for the other values of
$\mu$. $|\cdots,\eta\rangle^{(0)}$ denotes the Fock vacuum labelled by
the quantum numbers $\cdots$. In the untwisted sector $\cdots$
correspond to the quantum numbers $\{\bk,\bm\}$, whereas  in the
twisted sector $\cdots$ stand for the quantum numbers
$\{\bk,\bb\}$. The Fock vacuum in the twisted RR sector is degenerate,
but the state appearing in (\ref{bound}) is uniquely determined by the
condition coming from the fermionic zero modes in
Eq.~\refb{esx1}. $\NN$, $\wt\NN$ are normalisation factors to be
determined later. 

We are interested in the tree level amplitude that describes the
exchange of closed string states between two identical non-BPS
Dp-branes which are located at the same fixed  
point of $T^4/g$ (at $\bb=0$, say) and whose positions along the
non-compact directions (indexed by $\D_c$) are  described by $\ba_1$
and $\ba_2$.\footnote{In order to compute the self-interaction of such
a non-BPS D-brane we simply need to take $\ba_1=\ba_2$.} This
amplitude is given by 
\be
\int_{0}^\infty dl \, \langle \wt Dp,\ba_1,\bb=0,\eps |
e^{-l H_c} | \wt Dp,\ba_2,\bb=0,\eps\rangle
\,,
\ee
where $H_c$ is the closed string hamiltonian in light cone gauge,
\be
H_c = \pi \vec p^2 + {1\over 4\pi} \vec w^2 + 2\pi
\sum_{\mu=0,3,\ldots,9} \left[
\sum_{n=1}^{\infty} (\alpha^\mu_{-n} \alpha^\mu_{n}
      + \tilde{\alpha}^\mu_{-n} \tilde{\alpha}^\mu_{n})
+ \sum_{r>0} (\psi^\mu_{-r} \psi^\mu_r
      + \tilde\psi^\mu_{-r}\tilde\psi^\mu_{r} ) \right] + 2 \pi C_c
\,.
\ee
The constant $C_c$ takes the value $-1$ in the untwisted NSNS sector,
and $0$ in the untwisted RR, twisted NSNS, and twisted RR
sectors. $\vec p$ and $\vec w$ denote the momentum and winding charges
as usual. Using standard techniques this amplitude can be determined
to be 
\be\label{clos1}
{1\over 2}
\int_0^\infty dl\, l^{-{5-p \over 2}} e^{-{(\ba_1-\ba_2)^2 \over 4\pi l}}
\left[ {\cal N}^2 \left(\prod_{i=6}^{9} \sum_{m_i\in\Zop}
e^{-l\pi m_i^2/R_i^2}\right) {f_3^8(q) - f_4^8(q) \over f_1^8(q)}
- \widetilde{\cal N}^2 {f_2^4(q) f_3^4(q) \over f_1^4(q) f_4^4(q)}
\right]\, ,
\ee
where $q=e^{-2\pi l}$. In order to determine the normalisation
constants, we apply a modular transformation, setting $t=1/2l$, which
converts the closed string tree amplitude into an open string loop
amplitude. In the present case, using the transformation properties of
the $f_i$ functions,
\be\label{trans}
\begin{array}{rclrcl}
f_1(e^{-\pi/t}) & = & \sqrt{t} f_1(e^{-\pi t})\,, \qquad &
f_2(e^{-\pi/t}) & = & f_4(e^{-\pi t})\,, \\
f_3(e^{-\pi/t}) & = & f_3(e^{-\pi t})\,, \qquad &
f_4(e^{-\pi/t}) & = & f_2(e^{-\pi t})\,,
\end{array}
\ee
together with the identity
\be \label{ezzz1}
\sum_{m\in\Zop} e^{-\pi l (m/R)^2} = {R \over \sqrt{l}}
\sum_{n\in\Zop} e^{-2t\pi (nR)^2} \,,
\ee
we can express \refb{clos1} as
\ben\label{closamp1}
&& {1\over 2} \int_0^\infty {dt \over 2t} t^{-{p+1\over 2}}
\tilde{q}^{{(\ba_1-\ba_2)^2 \over 2 \pi^2}}
2^{5-p \over 2} \left[ 4 {\cal N}^2 \left(\prod_{j=6}^9 R_j \right)
\left(\prod_{i=6}^{9} \sum_{n_i\in\Zop} \tilde{q}^{2 n_i^2 R_i^2}
\right) 
{f_3^8(\tilde{q}) - f_2^8(\tilde{q}) \over f_1^8(\tilde{q})} \right.
\nonumber \\
&& \qquad\qquad\qquad\qquad\qquad\qquad\qquad \left.
- \widetilde{\cal N}^2 {f_4^4(\tilde{q}) f_3^4(\tilde{q}) \over
f_1^4(\tilde{q}) f_2^4(\tilde{q})} \right] \,,
\een
where $\tilde{q}=e^{-\pi t}$. This agrees with the result of the open
string calculation (\ref{ex10}) provided 
that
\be \label{enorma}
32 R_6 R_7 R_8 R_9 {\cal N}^2 = {A \over (2\pi)^{p+1}} \qquad \qquad
2 \widetilde{\cal N}^2 = {A \over (2\pi)^{p+1}} \,.
\ee
Using these values of $\NN$ and $\wt\NN$, (\ref{clos1}) reduces to  
\ben\label{closamp2}
& & {A\over (2\pi)^{p+1}} {1\over 4}
\int_0^\infty dl\, l^{-{5-p \over 2}} e^{-{(\ba_1-\ba_2)^2 \over 4\pi l}}
\left[ {1\over 16 R_6 R_7 R_8 R_9}
\left(\prod_{i=6}^{9} \sum_{m_i\in\Zop}
e^{-l\pi m_i^2/R_i^2} \right)
{f_3^8(q) - f_4^8(q) \over f_1^8(q)} \right. \nonumber \\
& & \qquad\qquad\qquad\qquad\qquad\qquad\qquad\qquad\qquad \left.
- {f_2^4(q) f_3^4(q) \over f_1^4(q) f_4^4(q)}
\right]\,.
\een
At the critical radii $R_i={1\over\sqrt{2}}$ for
$i=6,7,8,9$, the sums over $m_i$ can be simplified using
(\ref{ex14}), and together with (\ref{ex9a}) this implies that
the integrand in (\ref{closamp2}) vanishes identically.

{}From the point of view of the closed string calculation, the first
line in (\ref{closamp2}) contains the attractive (bulk) gravitational
force that is mediated by the massless fields from the untwisted NSNS
sector, whereas the second line contains the repulsive force between
objects carrying the same charge with respect to the twisted RR
sector. The latter is independent of the radii of the transverse
directions, but the former is inversely proportional to the radii, and
also contains winding contributions which become less and less
relevant as the radii are increased. Indeed, in the uncompactified 
theory, the untwisted sector contribution to the integrand would be
proportional to $l^{-(9-p)/2}$ for large $l$, indicating that the 
gravitational attraction is much weaker at long distances than the
repulsive force due to the charges.

\subsection{Example 2: Pair of BPS D-branes wrapped on supersymmetric
cycles of K3} \label{s22} 

This is the same example discussed in section \ref{s12}. In this case
we are interested in a system containing a BPS D-$(p+1)$-brane and its
anti-brane in type IIB/IIA on $T^4$ modded out by
$\II_4\cdot(-1)^{F_L}$. The branes stretch along the $x^9$ direction, 
and the other $p+1$ directions of their world-volumes are in the
non-compact directions. (In particular, they lie in the subspace
$x^i=b^i$ for $6\le i\le 8$ where $b^i$ can take values $0$ or 
$\pi R_i$.) None of the branes carry any Wilson line. If $\ba$ denotes
the location of such a brane in the non-compact directions
($\subset (x^1,\ldots, x^5)$) transverse to the brane, the boundary
state describing this BPS D-$(p+1)$ brane is of the form
\cite{9805019} 
\ben\label{Dpp} 
|D(p+1),\ba,\bb,\eps,\ko\rangle & = &
{1 \over 2} \Bigl(|B(p+1),\ba,\bb\rangle_{NSNS;U}
                      + \epsilon |B(p+1),\ba,\bb\rangle_{RR;U} \Bigr)
\nonumber \\
& & \quad + {1 \over 2 \sqrt{2}}
\epsilon\ko\Bigl(|B(p+1),\ba,\bb;0\rangle_{NSNS;T}
                 + \epsilon |B(p+1),\ba,\bb;0\rangle_{RR;T} \Bigr)
\nonumber \\
& & \quad + {1 \over 2 \sqrt{2}} \epsilon\ko\Bigl(
 |B(p+1),\ba,\bb;\pi R_9\rangle_{NSNS;T}
      \nonumber \\
& & \qquad\qquad\qquad 
+ \epsilon |B(p+1),\ba,\bb;\pi R_9\rangle_{RR;T} \Bigr)
\,. \nonumber \\
\een
$\eps$ and $\ko$ can take values $\pm 1$ and denote the sign of the
untwisted sector RR charge and the twisted sector RR charge,
respectively. (Thus $\epsilon=\pm$ corresponds to the brane and the
anti-brane, respectively.) The suffices indicate the closed string
sector to which each state belongs. For the case of the twisted
sectors, NSNS;T  and RR;T, we have also indicated whether the state is
in the sector localised at the fixed point $x^6=x^7=x^8=x^9=0$ of
$T^4/g$ (as is the case for the states in the second line of
Eq.~\refb{Dpp}) or at the fixed point $x^6=x^7=x^8=0$, $x^9=\pi R_9$
(as is the case for the states in the third and fourth line of
Eq.~\refb{Dpp}). Note that the sign of the twisted sector
charges at the $x^9=0$ and the $x^9=\pi R_9$ ends are taken to be the
same. This is a consequence of the fact that the branes do not carry
any Wilson line along $x^9$; more general cases have been discussed
in Ref.~\cite{9805019}. The different boundary states appearing on the
right hand side of Eqs.~\refb{Dpp} are given as
\ben \label{extra2}
|B(p+1),\ba,\bb\rangle_{NSNS;U}  & = &
{1 \over \sqrt{2}} \Bigl(|B(p+1),\ba,\bb,+\rangle_{NSNS;U} \nonumber
\\
& & \qquad\qquad    - |B(p+1),\ba,\bb,-\rangle_{NSNS;U} \Bigr)
\nonumber \\
|B(p+1),\ba,\bb\rangle_{RR;U}  & = &
{1 \over \sqrt{2}} \Bigl(|B(p+1),\ba,\bb,+\rangle_{RR;U}
                         + |B(p+1),\ba,\bb,-\rangle_{RR;U} \Bigr)
\nonumber \\
|B(p+1),\ba,\bb;c^9\rangle_{NSNS;T} & = &
{1 \over \sqrt{2}} \Bigl(|B(p+1),\ba,\bb,+;c^9\rangle_{NSNS;T}
       \nonumber
\\
& & \qquad\qquad     + |B(p+1),\ba,\bb,-;c^9\rangle_{NSNS;T} \Bigr)
\nonumber \\
|B(p+1),\ba,\bb;c^9\rangle_{RR;T} & = &
{1 \over \sqrt{2}} \Bigl(|B(p+1),\ba,\bb,+;c^9\rangle_{RR;T} \nonumber
\\
& & \qquad\qquad     + |B(p+1),\ba,\bb,-;c^9\rangle_{RR;T}
\Bigr)\,,\nonumber \\
\een
where
\ben \label{extra1}
|B(p+1),\ba,\bb,\eta\rangle_{NSNS;U}
& = & {\cal M} \int
\left(\prod_{\mu\in\D_c} dk^\mu e^{i \bk \cdot \ba}\right)
\left(\prod_{j=6}^{8} \sum_{m_j\in\Zop}
e^{i m_j b^j/R_j}\right) 
\nonumber \\
&& \qquad \sum_{n_9\in\Zop}
\wh{|B(p+1),\bk,\bm,n_9,\eta\rangle}_{NSNS;U} 
\nonumber \\
|B(p+1),\ba,\bb,\eta\rangle_{RR;U} & = & 4i {\cal M} \int
\left(\prod_{\mu\in\D_c} dk^\mu e^{i \bk \cdot \ba} \right)
\left(\prod_{j=6}^{8} \sum_{m_j\in\Zop}
e^{i m_j b^j/R_j}\right) 
\nonumber \\
&& \qquad \sum_{n_9\in\Zop}
\wh{|B(p+1),\bk,\bm,n_9,\eta\rangle}_{RR;U}  
\nonumber  \\
|B(p+1),\ba,\bb,\eta;c^9\rangle_{NSNS;T} & = & 2 \widetilde{\cal M}
\int \left(\prod_{\mu\in\D_c} dk^\mu e^{i \bk \cdot\ba} \right)
\wh{|B(p+1),\bk,\bb,\eta;c^9\rangle}_{NSNS;T} 
\nonumber
\\
|B(p+1),\ba,\bb,\eta;c^9\rangle_{RR;T} & = & 2 i \widetilde{\cal M}
\int \left(\prod_{\mu\in\D_c} dk^\mu e^{i \bk \cdot\ba} \right)
\wh{|B(p+1),\bk,\bb,\eta;c^9\rangle}_{RR;T}\,. 
\nonumber \\
\een
$c^9$ in Eq.~\refb{extra1} takes values 0 or $\pi R_9$, and corresponds
to twisted sector states associated to the fixed points $(\bb,0)$ and
$(\bb,\pi R_9)$, respectively. $n_9$ labels the winding number along
$x^9$, and $\D_c$ is defined as before. The coherent momentum (and
winding) eigenstates
$\wh{|B(p+1),\bk,\bm,n_9,\eta\rangle}$ 
(in the case
of
the untwisted sectors) and $\wh{|B(p+1),\bk,\bb,\eta;c^9\rangle}$ (in
the case of the twisted sectors) are again given by the same formula
as in (\ref{bound}), the only difference being that now $\C$ contains
also $\mu=9$ and $\hat\D$ does not. 
Furthermore the moding of the
fermions in the untwisted RR sector is opposite to that in the
untwisted NSNS sector, and that in the twisted NSNS 
sector is opposite to that in the twisted RR sector discussed below
\refb{bound}.  

The normalisation constants ${\cal M}$ and 
$\widetilde{\cal M}$ can be
determined from the tree level diagram that describes the exchange of
closed string states between two such branes. We take them to be
located at the same point in the compact space (say at $\bb=0$), but at
different points $\ba_1$ and $\ba_2$ in the non-compact space. We also
take them to carry the same twisted sector RR charges. The relevant
term describing 
the interaction between a pair of branes of this type is
\be
\int_{0}^\infty dl \, \langle D(p+1),\ba_1,\bb_1=0,\eps_1,\ko |
e^{-l H_c} | D(p+1),\ba_2,\bb_2=0,\eps_2,\ko\rangle 
\ee
which can be evaluated to give
\ben\label{cal1}
& & {1\over 4}
\int_{0}^{\infty} dl \, l^{-{5-p\over 2}}
e^{-{(\ba_1-\ba_2)^2 \over 4\pi l}}
\left[ {\cal M}^2 
\left(\sum_{n_9\in\Zop} e^{-l\pi R_9^2 n_9^2} \right)
\left(\prod_{i=6}^{8} \sum_{m_i\in\Zop} e^{-l\pi m_i^2/R_i^2}\right)
\right. \nonumber \\
& & \qquad\qquad\qquad\qquad\qquad\qquad\qquad \times
{f_3^8(q) - f_4^8(q) - \epsilon_1 \epsilon_2 f_2^8(q)
\over f_1^8(q)}  \nonumber \\
& & \qquad\qquad\qquad\qquad\qquad \left.
- \widetilde{\cal M}^2 
{f_2^4(q) f_3^4(q) \over f_1^4(q) f_4^4(q)}
(1- \epsilon_1 \epsilon_2)
\right]
\,.
\een
$\epsilon_j=\pm$ distinguishes between brane or anti-brane
for the branes localised at $\ba_j$, $j=1,2$. Again $q=e^{-2\pi l}$,
and the first two lines come from the untwisted NSNS and RR sector,
whereas the last line combines the contributions from the two twisted
NSNS and RR sectors at $x^9=0$ and $x^9=\pi R_9$.  Using (\ref{ex9a})
it is clear that the amplitude vanishes identically if the system
preserves supersymmetry, {\it i.e.} for $\epsilon_1=\epsilon_2$. 

We can rewrite the amplitude again in terms of open string coordinates
by setting $t=1/2l$. Then (\ref{cal1}) becomes
\ben\label{cal2}
& &
{1\over 4}
\int_{0}^{\infty} {dt \over 2t} \tilde{q}^{{(\ba_1-\ba_2)^2
\over 2\pi^2}}
t^{-{p+1 \over 2}} \left[ {\cal M}^2  2^{(9-p)/2}
{R_6 R_7 R_8 \over R_9}
\left(\sum_{m_9\in\Zop} \tilde{q}^{2 m_9^2/R_9^2}\right)
\right. \nonumber\\
& & \qquad\qquad\qquad\qquad\qquad\qquad\qquad
\times \left(\prod_{i=6}^{8} \sum_{n_i\in\Zop}
\tilde{q}^{2 n_i^2 R_i^2} \right)
 {f_3^8(\tilde{q}) - f_2^8(\tilde{q})
- \epsilon_1 \epsilon_2 f_4^8(\tilde{q})
\over f_1^8(\tilde{q})} 
\nonumber \\
& & \qquad\qquad\qquad\qquad\qquad \left.
- \widetilde{\cal M}^2  2^{(5-p)/2}
{f_4^4(\tilde{q}) f_3^4(\tilde{q}) \over
f_1^4(\tilde{q}) f_2^4(\tilde{q})}
(1- \epsilon_1 \epsilon_2)
\right]
\,,
\een
where we have again used (\ref{trans}) and \refb{ezzz1}. For
$\eps_1=-\eps_2$ this agrees with \refb{ex22} provided that 
\be \label{enorma2}
32 {R_6 R_7 R_8 \over R_9} {\cal M}^2 = {A \over (2\pi)^{p+1}}\,,
\qquad\qquad 2 \widetilde{\cal M}^2 = {A \over (2\pi)^{p+1}}\,.
\ee
Now that we have determined the constants, (\ref{cal1}) becomes
\ben\label{cal3}
& & {1\over 8} {A \over (2\pi)^{p+1}}
\int_{0}^{\infty} dl\, l^{-{5-p \over 2}}
e^{-{(\ba_1-\ba_2)^2  \over 4\pi l}}
\left[ {1\over 16} {R_9 \over R_6 R_7 R_8} 
\left(\sum_{n_9\in\Zop} e^{-l\pi R_9^2 n_9^2}\right)
\left(\prod_{i=6}^{8} \sum_{m_i\in\Zop} e^{-l\pi m_i^2/R_i^2}\right)
\right. \nonumber \\
& & \qquad\qquad\quad
\left. {f_3^8(q) - f_4^8(q) - \epsilon_1 \epsilon_2 f_2^8(q)
\over f_1^8(q)} 
- {f_2^4(q) f_3^4(q) \over f_1^4(q) f_4^4(q)}
(1- \epsilon_1 \epsilon_2)
\right]
\,.
\een
For $\epsilon_1=-\epsilon_2$, using (\ref{ex9a}), the integrand
simplifies to
\ben\label{cal4}
& & {1\over 4} {A \over (2\pi)^{p+1}}
l^{-{5-p \over 2}}
e^{-{(\ba_1-\ba_2)^2 \over 4\pi l}}
\left[ {R_9 \over 16 R_6 R_7 R_8} 
\left(\sum_{n_9\in\Zop} e^{-l\pi R_9^2 n_9^2}\right)
\left(\prod_{i=6}^{8} \sum_{m_i\in\Zop} e^{-l\pi m_i^2/R_i^2}\right)
{f_2^8(q) \over f_1^8(q)}  \right. \nonumber \\
& & \qquad\qquad\qquad\qquad\qquad\qquad\qquad\left.
- {f_2^4(q) f_3^4(q) \over f_1^4(q) f_4^4(q)}
\right]
\,.
\een
At the critical radius, $R_6=R_7=R_8=1/\sqrt{2}$,
$R_9=\sqrt{2}$, the momentum and winding sums simplify as before using
\refb{ex14},
and (\ref{cal4}) vanishes.

{}From the point of view of the closed string calculation, the first
term in the integrand of (\ref{cal3}) contains the gravitational
interaction and the interaction due to the exchange of the untwisted
RR fields. In the case of the brane-anti-brane pair, the two branes
have opposite charge with respect to the untwisted RR fields, and
therefore both interactions are attractive.
Indeed, for $\epsilon_1=-\epsilon_2$ the last factor in the first term
in \refb{cal3} is $2 f_2^8/f_1^8$, and so the first term is strictly
positive. The second term contains the interaction due to the states
in the twisted sectors. For the case of brane-anti-brane pair (without
Wilson line) the interactions at both fixed points of $T^4/g$ 
are equal and repulsive. This is due to the fact that the two states
under consideration carry the opposite twisted NSNS and the same
twisted RR charge at each fixed point; both lead to a repulsive
force. At the critical point the attractive interaction due to the
untwisted sector fields cancels the repulsive interaction due to the
twisted sector fields at all distance scales.

\subsection{Example 3: A pair of BPS D-branes wrapped on
non-super\-sym\-met\-ric cycles of K3} 

The theory under consideration is again type IIB/IIA on $T^4$ modded
out by $\II_4\cdot(-1)^{F_L}$. We shall take a pair of non-BPS
D$p$-branes of the type described in section \ref{s21}, one at
$(\ba,\bb$=$0)$ and the other at $(\ba, \bb$=$(0,0,0,\pi R_9))$. Note
that we have taken the locations of the two branes in the non-compact
directions to be identical. In the dual type IIA/IIB string theory on
$T^4/\II_4$, this describes a system containing a pair of non-BPS
branes, corresponding to BPS D-branes wrapped on two homologically
distinct non-supersymmetric 2-cycles, situated at the same location in
the non-compact space-time. The boundary state describing this system
is given by:
\be \label{e23one}
|\wt Dp_1,\wt Dp_2\rangle = |\wt Dp, \ba, \bb=0, \eps_1\rangle
+|\wt Dp, \ba, \bb=(0,0,0,\pi R_9), \eps_2\rangle\, ,
\ee
where the boundary states appearing on the right hand side of this
equation are identical to the ones defined in section \ref{s21}.
The amplitude of interest is
\be \label{e23two}
\int_0^\infty dl\, \langle \wt Dp_1,\wt Dp_2|
e^{-lH_c}| \wt Dp_1, \wt Dp_2\rangle\, .
\ee
This can be computed easily using the expression for the boundary states
given in section \ref{s21}, Eqs.~\refb{Dp}$-$\refb{bound}, and is
given by 
\ben \label{e23three}
{1\over 2}
\int_0^\infty dl& l^{-{5-p \over 2}}
& \Bigl[ 4{\cal N}^2 \Big(\prod_{i=6}^{8} \sum_{m_i\in\Zop}
e^{-l\pi m_i^2/R_i^2}\Big)
\Big(\sum_{m_9\in 2\Zop} e^{-l\pi m_9^2/R_9^2}\Big)
{f_3^8(q) - f_4^8(q) \over
f_1^8(q)} \nonumber \\
&&  \quad
- 2\widetilde{\cal N}^2 {f_2^4(q) f_3^4(q) \over f_1^4(q) f_4^4(q)}
\Bigr]\, ,
\een
where $\NN$ and $\wt\NN$ have been defined in \refb{enorma}. Using
these, \refb{e23three} can be rewritten as
\ben \label{e23four}
&& {1\over 2}
{A\over (2\pi)^{p+1}}\int_0^\infty dl\, l^{-{5-p \over 2}}
\nonumber \\
&& \left[ {1\over 8R_6R_7R_8R_9} \Big(\prod_{i=6}^{8}
\sum_{m_i\in\Zop} 
e^{-l\pi m_i^2/R_i^2}\Big)
\Big(\sum_{m_9\in 2\Zop} e^{-l\pi m_9^2/R_9^2}\Big)
{f_3^8(q) - f_4^8(q) \over f_1^8(q)}
- {f_2^4(q) f_3^4(q) \over f_1^4(q) f_4^4(q)}
\right]\, . \nonumber \\
\een
By the same identities used in section \ref{s21} we see that this
vanishes exactly at
\be \label{e23five}
R_6=R_7=R_8={1\over \sqrt 2}, \quad R_9=\sqrt 2\, . 
\ee
Thus at this critical point the spectrum of open string states on this
system develops exact Bose-Fermi degeneracy.

Note that at this critical radius there is an exact marginal
deformation which takes this system to the system discussed in section
\ref{s22} \cite{9805019,9808141}. Thus it is natural to ask if the
bose-fermi degeneracy of the spectrum survives all along this line of
marginal deformation. This is however not the case. From the arguments
in Ref.~\cite{9808141} it is easy to see that the spectrum in the
Ramond sector does not change during this marginal deformation,
whereas the spectrum in the NS sector certainly does. Thus the
bose-fermi degeneracy in the spectrum can only appear at special
points along this critical line.

\subsection{Example 4: A pair of BPS D-branes wrapped on
supersymmetric cycles of K3} \label{s24} 

This example will involve a system similar to that discussed in
section \ref{s22}, $-$ the only difference being that instead of
taking a pair of anti-parallel BPS D$(p+1)$ branes in type IIB/IIA on
$T^4$ modded out by $(-1)^{F_L}\cdot \II_4$, we shall consider a pair
of `orthogonal' BPS D$(p+1)$ branes.\footnote{Some aspects of tachyon
condensation on D-branes at angles have been recently discussed in
\cite{9905080}.} In particular we shall take both D-branes to span the
same non-compact directions, but take the first D-brane to lie along
$x^9$ at $\bb\equiv(x^6=0, x^7=0, x^8=0)$, and the second D-brane to
lie along $x^8$ at $\bb'\equiv(x^6=0,x^7=0,x^9=0)$.  If $\ko$ denotes
the sign of the twisted RR charge of the first D-brane at the fixed
points $(0,0,0,0)$ and $(0,0,0,\pi R_9)$,
and ${\ko}'$ denotes the sign of the twisted RR charge of the second
D-brane at the fixed points $(0,0,0,0)$ and $(0,0,\pi R_8,0)$, $\ba$
and  $\ba'$ denote their locations in the non-compact directions, and 
$\eps$, $\eps'$ denote the sign of the untwisted sector RR charges,
then the boundary state of the combined system is given by 
\be\label{e24one} 
|D(p+1),\ba,\bb,\eps,\ko\rangle + |D(p+1),\ba',\bb',\eps',\kop\rangle'
\, , \ee where $|D(p+1),\ba,\bb,\eps,\ko\rangle$ is the boundary state
defined in section \ref{s22}, and
$|D(p+1),\ba',\bb',\eps',\kop\rangle'$ is related to the boundary
state defined in section \ref{s22}\ by exchanging the 8th and the 9th
coordinates everywhere.

In computing the amplitude describing the emission and reabsorption of
closed strings from this system, we note that since each system is
individually BPS the amplitude for emission and reabsorption of a
closed string by the same D-brane will vanish identically. Thus the
relevant amplitude which needs to be analysed is 
\be \label{e24two}
\int_0^\infty dl \,
\langle D(p+1),\ba,\bb,\eps,\ko | e^{-l H_c} | 
D(p+1),\ba',\bb',\eps',\kop\rangle' \, .
\ee
Each of the boundary states in this expression has components in the
untwisted NSNS and RR sectors, as well as the twisted NSNS and RR
sectors. The calculation can be simplified by  noting that in any 
sector which contains zero modes of the fermion fields
$\psi^8,\psi^9,\wt\psi^8,\wt\psi^9$, the two Fock vacua appearing in
the expression for the two boundary states in Eq.~\refb{e24two} are 
orthogonal since they satisfy different constraints coming from the
fermionic zero modes in the analogue of Eq.~\refb{esx1}. Thus the
contribution to \refb{e24two} from these sectors vanishes. This leaves
us with the untwisted NSNS and the twisted RR sectors. Furthermore,
among the various twisted sector RR states appearing in \refb{e24one},
only the sector that is localised at $x^6=x^7=x^8=x^9=0$ contributes
to \refb{e24two}, since this is the only sector that is shared by
both boundary states. The contribution can be easily evaluated, and
gives 
\ben \label{e24three}
&& {1 \over 4} \int dl \, e^{-{(\ba-\ba')^2\over 4\pi l}}
l^{-{5-p\over 2}}
\Biggl[ 2 {\cal M} {\cal M}'
\Bigl(\sum_{m_6,m_7\in \Zop} e^{-l\pi ( (m_6/R_6)^2 + (m_7/R_7)^2 )}
\Bigr) \nonumber \\
& & \quad \times 
{f_3(q)^6 f_4(q)^2 - f_4(q)^6 f_3(q)^2 \over f_1(q)^6 f_2(q)^2}
- {1 \over 2} \ko{\ko}'\widetilde{\cal M} \widetilde{\cal M}'
{f_2(q)^4 f_3(q)^2 f_4(q)^2  
\over f_1(q)^4 f_4(q)^2 f_3(q)^2} \Biggr]
\,,
\een
where $\MM$, $\wt\MM$ are given as in \refb{enorma2}, and $\MM'$,
$\wt\MM'$ are obtained from Eq.~\refb{enorma2} by exchanging
$8\leftrightarrow 9$:
\be \label{enorma3}
{\cal M}^2 = {A \over (2 \pi)^{p+1}} {R_9 \over 32 R_6 R_7 R_8}
\qquad\qquad
{{\cal M}'}^2 = {A \over (2 \pi)^{p+1}} {R_8 \over 32 R_6 R_7 R_9}\, ,
\ee
and
\be \label{enorma4}
\widetilde{{\cal M}}^2 = \widetilde{{\cal M}'}^2 =
{A \over 2 (2 \pi)^{p+1}}\, .
\ee
The integral \refb{e24three} then becomes
\ben \label{e24four}
I & = & {1 \over 16} {A \over (2\pi)^{p+1}}
\int_0^\infty dl \, e^{-{(\ba-\ba')^2\over 4\pi l}}
l^{-{5-p\over 2}}  \nonumber \\
&& \left[ {1 \over 4 R_6 R_7} \Bigl(
\sum_{m_6,m_7} e^{-l\pi ( (m_6/R_6)^2 + (m_7/R_7)^2 )}\Bigr) 
{f_3(q)^6 f_4(q)^2 - f_4(q)^6 f_3(q)^2 \over f_1(q)^6 f_2(q)^2}
- \ko{\ko}'{f_2(q)^4 \over f_1(q)^4 } \right] \,. \nonumber \\
\een
At the `critical radius' $R_6=R_7=1/2$, the sums over $m_6,m_7$ give
\be \label{e24fiveone}
\sum_{m_6,m_7} e^{-l\pi ( (m_6/R_6)^2 + (m_7/R_7)^2 )} =
\left(\sum_n q^{2n^2}\right)^2 \,.
\ee
This can be re-expressed in terms of the Jacobi $\vartheta$-function
$\vartheta_3(0|\tau)$ that we considered before in Eq.~(\ref{ex12}),
and the  Jacobi $\vartheta$-function $\vartheta_4(0|\tau)$, whose sum
and product representation is \cite{EMOT53}
\be \label{theta4}
\vartheta_4(0|\tau) = \sum_{n\in \Zop}  q^{n^2} (-1)^n
= \prod_{n=1}^\infty (1 - q^{2n}) (1 + q^{2n-1})^2
= f_1(q) f_4^2(q)\, .
\ee
Indeed, using (\ref{ex12}) and (\ref{theta4}), we find
\begin{eqnarray}
\vartheta_3^2(0|\tau)+ \vartheta_4^2(0|\tau) & = & 
\sum_{n,l} q^{n^2+l^2} + \sum_{n,l} (-1)^{n+l} q^{n^2+l^2} \nonumber
\\
& = & \sum_{n,l} (1+(-1)^{n+l}) \,
q^{{1\over 2}\left[(n+l)^2 + (n-l)^2\right]}
\nonumber \\
& = & 2 \sum_{r,s\, even} 
q^{{1\over 2}\left[r^2 + s^2\right]} 
\nonumber \\
& = & 2 \left(\sum_m q^{2 m^2} \right)^2 \,,
\end{eqnarray}
where we have set $r=n+l$ and $s=n-l$, and observed that if $r=n+l$ is
even, then so is $s=n-l$. Thus (\ref{e24fiveone}) can be rewritten as 
\be \label{e24five}
\left(\sum_n q^{2n^2}\right)^2 = {1 \over 2} f_1(q)^2 (f_3(q)^4 +
f_4(q)^4)\, ,
\ee
and the integrand in (\ref{e24four}) becomes, apart from the overall
factor of ${1\over 16} {A \over (2\pi)^{p+1}} e^{-(\ba-\ba')^2\over 4\pi l} 
l^{-{5-p\over 2}}$ and for $\ko{\ko}'=1$, 
\ben \label{e24six}
&& \left[ {1 \over 2}
{f_3(q)^2 f_4(q)^2 (f_3(q)^4 - f_4(q)^4) (f_3(q)^4 + f_4(q)^4) \over
f_1(q)^4 f_2(q)^2}
- {f_2(q)^4 \over f_1(q)^4 } \right]\nonumber \\
& & = 
{1 \over f_1(q)^4 f_2(q)^2} \left[{1 \over 2} f_3(q)^2 f_4(q)^2
(f_3(q)^8 - f_4(q)^8) - f_2(q)^6 \right]  \nonumber \\
& & = 
{f_2(q)^6 \over f_1(q)^4 f_2(q)^2} \left[{1 \over 2} (f_3(q)^2 f_4(q)^2
f_2(q)^2)
- 1 \right] = 0 \,,
\een
where we have used the identity \refb{ex13} in the last line.

The radius $R_6=R_7=1/2$ is critical in the sense that for $R_i<1/2$, 
$i=6,7$, the open string that begins on one D$(p+1)$-brane and ends on
the other contains a tachyon. Indeed, in this case the $C_o$ in
Eq.~\refb{ex2} is equal to $-(1/4)$, and 
the energy of the winding states with winding number $(n_6,n_7)$
along the 6th and the 7th directions is given by 
\be \label{e24seven}
- {\pi\over 4} + \pi (n_6 R_6)^2 + \pi (n_7 R_7)^2 \,.
\ee
The $n_6=n_7=0$ mode is projected out, but appropriate linear
combinations of the states with
$(n_6,n_7)=(\pm 1,0)$ or $(n_6,n_7)=(0,\pm 1)$ survive. At least one
of them is tachyonic whenever $R_6$ or $R_7$ is below $(1/2)$, and
the system is therefore unstable in this regime. For generic values of
the radii $R_8$ and $R_9$ there does not seem to be any simple system
of 
D-branes into which this system can decay. This leads us to suspect that
below the critical values of $R_6$ and/or $R_7$ the system forms a bound
state which cannot be described by a solvable boundary conformal field
theory of a system of D-branes.  

\subsection*{Acknowledgement}
This work was begun during the Extended Workshop on String Theory that
was held at The Abdus Salam International Centre for Theoretical
Physics, Trieste in June-July 1999. We would like to thank the ICTP for
hospitality during this workshop. 

\noindent M.R.G. is supported by a College Lectureship of Fitzwilliam
College, Cambridge.

\vfill\eject

\end{document}